\pdfoutput=1

\documentclass{IEEEtran}
\IEEEoverridecommandlockouts
\usepackage{cite}
\usepackage{amsmath,amssymb,amsfonts}
\usepackage{algorithmic}
\usepackage{graphicx}
\usepackage{textcomp}
\usepackage{xcolor}

\def\BibTeX{{\rm B\kern-.05em{\sc i\kern-.025em b}\kern-.08em
    T\kern-.1667em\lower.7ex\hbox{E}\kern-.125emX}}
\begin{document}

\title{An Auto Encoder For Audio Dolphin Communication}

\author{\IEEEauthorblockN{1\textsuperscript{st} Daniel Kohlsdorf}
\IEEEauthorblockA{\textit{Freelance Data Scientist} \\
Bremen, Germany \\
dkohlsdorf@gmail.com}
\and
\IEEEauthorblockN{2\textsuperscript{nd} Denise Herzing}
\IEEEauthorblockA{\textit{Wild Dolphin Project} \\
West Palm Beach, USA \\
dlherzing@wilddolphinproject.org}
\and
\IEEEauthorblockN{3\textsuperscript{rd} Thad Starner}
\IEEEauthorblockA{\textit{College of Computing} \\
\textit{Georgia Institute of Technology}\\
Atlanta, USA \\
thad@cc.gatech.edu }
}
\maketitle

\begin{abstract}
  Research in dolphin communication and cognition requires
  detailed inspection of audible dolphin signals. The manual
  analysis of these signals is cumbersome and time-consuming.
  We seek to automate parts of the analysis using modern deep learning
  methods. We propose to learn an autoencoder constructed from
  convolutional and recurrent layers trained in an unsupervised fashion.
  The resulting model embeds patterns in audible dolphin communication.
  In several experiments, we show that the embeddings can be used for
  clustering as well as signal detection and signal type classification.
\end{abstract}


\section*{Introduction}
\begin{figure}[ht]
  \centering
  \includegraphics[width=0.5\textwidth]{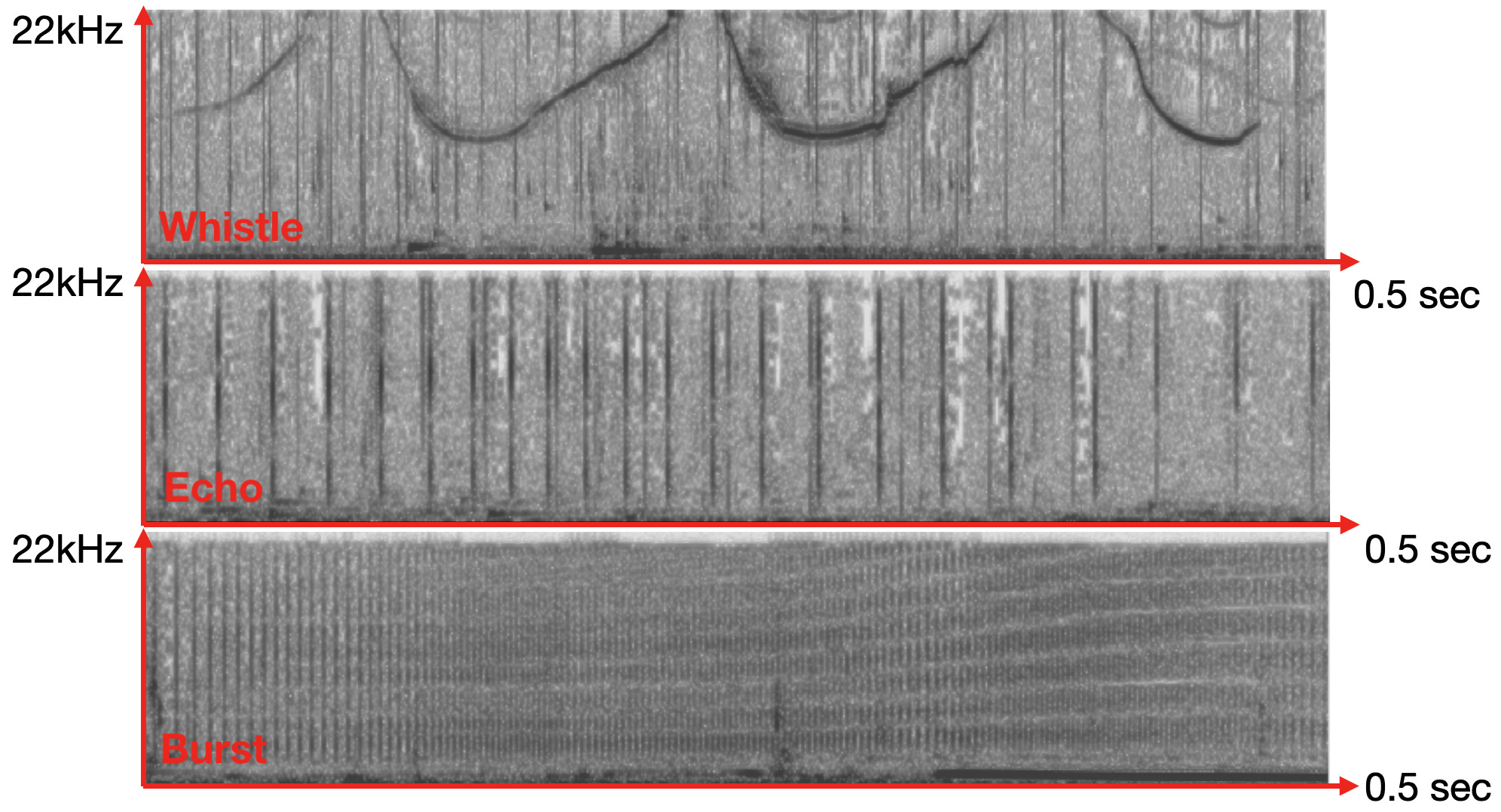}
  \caption{Example spectrogram of several dolphin signals. Top: a whistle, middle: a series of echolocation clicks,
  bottom: a burst pulse.}
  \label{fig:examples}
\end{figure}

Audible dolphin signals provide insight into dolphin cognition and social structure.
Marine mammalogists collect large datasets of underwater recordings when encountering
dolphins in the wild. For example, for 29 summers, each with 100 field days, researchers of the Wild Dolphin Project have collected audio and video data while 
observing wild Atlantic spotted dolphins (\emph{Stenella frontalis}) underwater, in the Bahamas. The analysis of this data involves
annotating the videos with observed dolphin behavior as well as dolphin names and audible signal
type categories. In order to understand dolphin communication, researchers desire to correlate patterns
in the audio with observed behavior. However, finding patterns in audible communication manually
involves intensive measurements and comparisons across multiple spectrograms of the field recordings. 
Every hour of field recordings requires ten hours of manual analysis.
We seek to automate several parts of the analysis:
signal detection, sound type classification, and pattern identification.
In the signal detection step, we locate dolphin signals temporally in the field recordings.
The sound type classification step is needed to automatically determine the type of dolphin signal.
There are several sound types in dolphin communication. Three prominent types are whistles, burst pulses
and echolocation (see Fig. \ref{fig:examples}) \cite{b13}. These types are often indicative of dolphin behavior.
For example, dolphins use signature whistles to name each other while echolocation is often used during
foraging.

Signal detection and signal type classification can help researchers to browse the data more
efficiently. However, the ultimate goal is to automatically identify patterns in audible dolphin communication.
A model of the patterns in dolphin signals needs to be robust to frequency shifting and time warping.
One source of these variations is that dolphins are known to shift their communication into higher frequency bands depending
on the noise floor. Other variations occur due to the angle of recording and the dolphins' sound production itself, among others.
Another challenge is that there is no fully annotated large dataset that could enable us to train a deep neural network.

We developed a deep autoencoder constructed from convolutional and recurrent layers.
Through several experiments we show that we can indeed train a model in an unsupervised manner that enables
us to find patterns in dolphin communication and can be used in a transfer learning setup.

Our contributions are

\begin{itemize}
\item A deep neural network for audible dolphin signals trained in an unsupervised manner
\item A clustering experiment using embeddings from our architecture
\item A dolphin signal detection experiment
\item A dolphin type classification experiment
\item Clustering all non-silent regions of a complete year of audio recordings.
\end{itemize}

\section*{Related Work}
Marine mammalogists use interactive tools for the manual analysis of animal communication recordings.
For example, Cornell's RAVEN \cite{b22} allows researchers to 
annotate animal audio recordings in spectrogram form.
Furthermore, Raven includes algorithms for animal signal detection in larger recordings. Noldus Observer
enables users to annotate audio and video data together. Observer is primarily used to annotate 
annotate animal behavior data manually and does not offer automatic analysis capabilities.
The automatic analysis of dolphin communication can increase the research speed of marine mammalogists.
Previous models required several separate machine learning algorithms. For example, one approach involves
learning a feature space that is invariant to signal shifts in frequency using convolutional k-means \cite{b3}.
The next step is to cluster all examples using the dynamic time warping distance in order to account for temporal warpings of the signal.
The last step is to learn a probabilistic model using a mixture of hidden Markov models. In order to
build a classifier on top of the unsupervised results, the algorithm uses a bag-of-words approach
on the cluster IDs \cite{b1, b2, b3}. One drawback is that these methods have to be tuned independently,
which requires significant manual labor.

Previously, researchers have used the above methods in isolation. For example, Lampert and O`Keefe
detect dolphin whistles in spectrograms under various distortions using hidden Markov models \cite{b8}.
The dynamic time warping distance is also used to measure similarity between
manually extracted contours from spectrograms of dolphin whistles \cite{b9}. 
Other models for dolphin communication classification and clustering use neural networks \cite{b10}.

There are also several specialized models for dolphin whistles. Two closely related methods are a Viterbi algorithm based pitch tracker \cite{b18}
and a Kalman filtering approach \cite{b20}. Other approaches include a frame-based Bayesian approach \cite{b19} and a pitch detection algorithm
designed for human telephone speech that 
is also capable of extracting whale vocalizations \cite{b21}. All of these approaches seek to extract the contour
of a dolphin whistle. While these methods are very effective for whistle based communication, we wish to model a wider variety of dolphin signals
and underwater artifacts.

In the proposed architecture we want to enable an end-to-end deep learning approach to the feature extraction step that handles
frequency shifts and temporal warps in one model. Furthermore, each pattern should be encoded by the model into
a single vector. Our model is inspired by the machine translation community. Specifically, we are inspired by
the encoder-decoder model \cite{b6}. 

In a survey \cite{b24} of neural networks
as models for animal communication, the authors suggest to use a sequence to sequence model on top of convolutional neural networks and show how these models can be used
for visualization. We will use a similar model and study its performance
on our dolphin communication dataset.

The encoder in these models encodes a sequence of words into a single vector using a many-to-one recurrent neural network. The decoder creates the translation from the embedding vector using a one-to-many
recurrent neural network. Instead of words, we pass the encoder the output of a convolutional layer, followed by max-pooling.
The decoder’s output is not the result itself, but the resulting sequence is passed to a deconvolution layer, similar to a convolutional autoencoder \cite{b7}. 
Instead of training the model for translation, we train the encoder and decoder
to reconstruct a spectrogram window of dolphin communication. In the following section, we describe our model in more detail.

\section*{Proposed Architecture}
Our goal is to learn a feature space that maps small spectrogram windows ($128$ spectrogram
frames or $3/4$ seconds) into a single embedding vector.
We propose an autoencoder to account for
signal variations in time and frequency. The first layers
of our encoder seek to achieve invariance to signal shifts in frequency by
using a convolutional layer followed by a pooling operation.
The recurrent layers compensate for time warping effects in the dolphin signals.
After encoding a window of dolphin communication into a single vector, the decoder reconstructs
the input window. Therefore, the encoding process is reversed. In other words, we
reconstruct a sequence with the length of the input sequence from the
embedding vector using several recurrent layers, and the signal is reconstructed by a series of deconvolutions.

\begin{figure}[ht]
  \centering
  \includegraphics[width=0.5\textwidth]{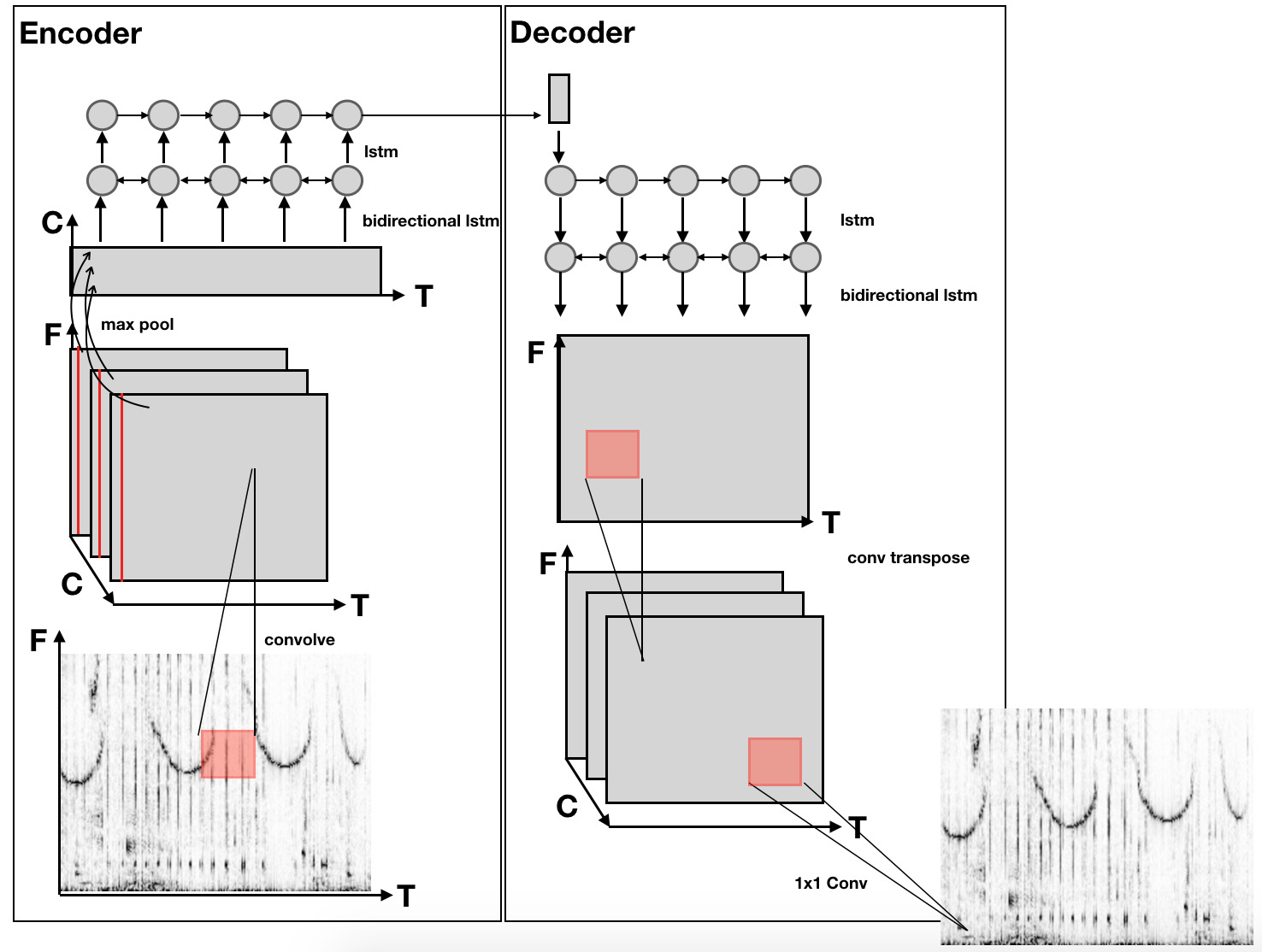}
  \caption{The autoencoding architecture. Left: the encoder constructed from
    convolutional and recurrent layers. Right: The decoder constructed from deconvolutional
    and recurrent layers. }
  \label{fig:architecure}
\end{figure}

Our architecture is shown in Fig. \ref{fig:architecure}, 
and our Keras implementation can be found at https://github.com/dkohlsdorf/wdp-ds/tree/v4.0/ml\_pipeline
\subsection*{Architecture Details}
We extract sliding windows

$$x=\{x_1 ... x_T\}, x_i \in \mathbb{R}^{F}$$

from a spectrogram with $F$ frequency bins. Each spectrogram window is convolved by $256$ filters resulting in a novel sequence:

$$\hat{x} = \{\hat{x_1} ... \hat{x_T}\}, \hat{x_i} \in \mathbb{R}^{F x 256}$$.

Each sample in the new sequence $\hat{x}$ has $256$ channels per frequency bin.
Each filter spans $0.02$ seconds and $680$ Hz. In the next step, we apply max pooling.
For every channel in a sample, we pool across all frequency bins.  In other words,
for each spectrogram frame and filter, we select the maximum response across all frequencies.
The result is a new sequence:

$$pooled(\hat{x}) = \{pooled(\hat{x_1}) ... pooled(\hat{x_T})\}, pooled(\hat{x_i}) \in \mathbb{R}^{256}$$

with the same length as the spectrogram and one dimension per filter. In this way, we account for frequency shifts
since we discard the frequency of the maximum response of a filter.

In order to account for temporal warping effects, we use the filter
response sequence as input to several layers of recurrent units.
The first layer is a many-to-many bidirectional long short term memory (LSTM) cell \cite{b12, b15}.
The second layer is a many-to-one LSTM. The resulting vector is our embedding:

$$e = LSTM(pooled(\hat{x})), e \in R^{128}$$ 

Dolphin signals can develop in complex ways over time. Therefore, we decided to use LSTM cells
in our recurrent neural network for their ability
to hold information longer. The first recurrent layer returns a sequence
of the same length as the original sequence (many-to-many) serving as the input to the final
embedding layer. Furthermore, the first recurrent layer is a bidirectional LSTM enabling
the outputs to depend on information from the future and the past. The embedding layer
is a simple LSTM with one output (many-to-one).

The decoder aims to reconstruct the input spectrogram from
the encoding by reversing the encoding process.
First, we build a new sequence from the embedding vector using
a one-to-many LSTM which is followed by a bi-directional LSTM.
The output is a sequence of the same length as the spectrogram:

$$x'= \{x'_1 ... x'_T\}, x'_i \in R^{F}$$.

The hope is that the embedding vector holds enough information to construct
a whole sequence resembling the spectrogram input from it. We then apply a convolution
with 256 filters:

$$\hat{x'}= \{\hat{x'_1} ... \hat{x'_T}\}, \hat{x'_i} \in R^{F x 256}$$.

followed by a 1x1 deconvolution which creates the final spectrogram.

Using the convolutions in the decoder we upsample the number of channels to model the decoder as a reverse of the encoder. 
In order to reshape the final sequence to the same shape as the input spectrogram, we apply a 1x1 deconvolution layer
with one kernel. This technique reduces the shape of each sample from $R^{F x 256}$ to $R^{F}$.

\subsection*{Preprocessing And Training}
Before the training, we compute the spectrogram using a $0.01$ second window
with a $0.005$ second skip. Then, we apply a Hanning window before
computing the discrete Fourier transform. We normalize each of the spectrogram
windows ($3/4$ seconds) to its standard score. Therefore, we compute the mean
and standard deviation for each spectrogram frame and then subtract the mean of
the frame and divide it by its standard deviation.

During training, we use a batch size of $50$ spectrogram windows and train for $128$ epochs.
We optimize the model by minimizing the mean square error between the input spectrogram
and the decoder's reconstruction using the ADAM optimizer \cite{b16}.

\subsection*{Transfer Learning}
The autoencoder can be trained in an unsupervised manner.
Collecting a large amount of unlabeled data is easy
and can be used to learn a feature space embedding appropriate for dolphin communication.
Once the model is trained, we can use a smaller labeled dataset
to construct a classifier on top of the embedding.

In our experiments, we will show how to use the embedder to distinguish
between spectrogram windows with dolphin communication
and water noise and how to use the embedder to distinguish between
different dolphin communication types.

When building a classifier we simply add three dense layers on top of
the encoder's output and freeze all other layers' weights except for the last LSTM.
The first two dense layers' activations are rectified linear units and
the last layer's activation is either a sigmoid for binary classification
problems or a softmax activation for multiclass problems.
By retraining on a labeled dataset we can easily construct a classifier.

\section*{Experiments}
With the following experiments, we will highlight several aspects
of our model. We run a clustering experiment to show that similar
patterns are embedded close to each other. In two experiments
using the same model, we show that the encoder can be adopted
to tasks such as dolphin signal detection and dolphin type classification.

\subsection*{Datasets}
In our experiments we use data from in-water field recordings collected by the Wild Dolphin Project.
The audio data is extracted from video files from underwater cameras filming Atlantic spotted dolphins.

For our experiments we collect four datasets:
\begin{itemize}
\item An unlabeled dataset of 24 minutes collected form several field recordings from 2008, 2010 and 2012
\item A small signal detection dataset containing 33 seconds of
signal and 83 seconds of noise from field recordings in 2011
\item A full year of field recordings consisting of ̃16 hours collected in 2011.
\item A small labeled signal-type classification dataset
\end{itemize}

The classification dataset is labeled for dolphin signals containing:
\begin{itemize}
\item Noise: basic water noise, 82 seconds
\item Echolocation: echolocation of dolphins, 118 seconds
\item Burst Pulses: dense click packages, 71 seconds
\item Whistles: the dolphins’ whistles, 64 seconds
\end{itemize}

The last dataset is a full year of field recordings. It consists
of $\approx 16$ hours collected in 2011. All audio is recorded with a
sample rate of 44100Hz.

\subsection*{Autoencoding}
\begin{figure}[ht]
  \centering
  \includegraphics[width=0.5\textwidth]{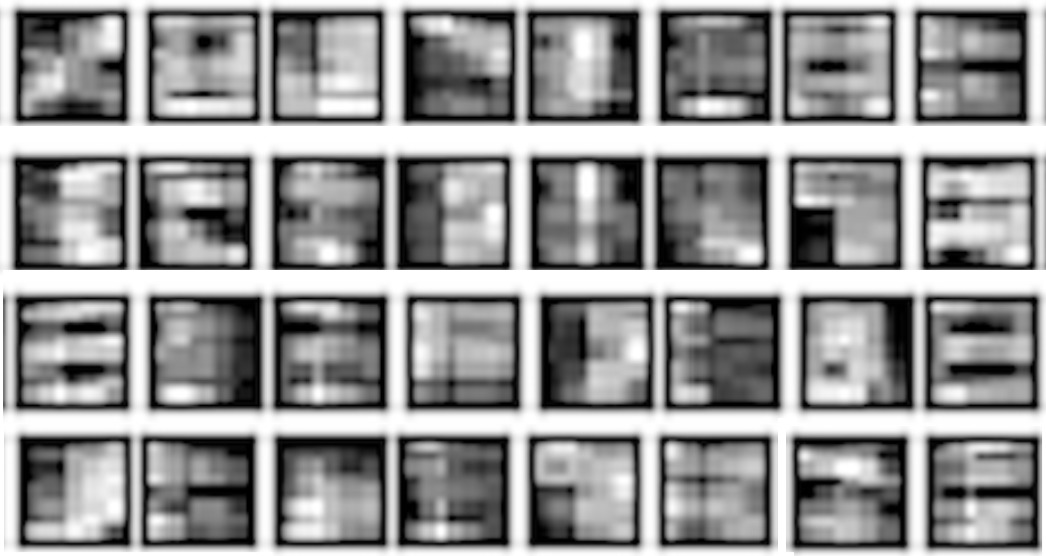}
  \caption{A subset of the first layer convolutional kernels. Kernels that show multiple lines
    will be able to detect burst pulse signals while kernels with a single line
    will detect  dolphin whistles. Vertical lines in a kernel indicate the presence
    of echolocation clicks.
  }
  \label{fig:embedder_filters}
\end{figure}

In the first experiment, we train our autoencoder on the $22$ minute long unsupervised dataset.
In order to inspect the learned model, we visualize the first layer's convolutional kernels.
As one can see in Fig. \ref{fig:embedder_filters} the result is filters representing
up and down sweep in whistles (single line) as well as several representing burst pulses (multiple lines).

\begin{figure}[ht]
  \centering
  \includegraphics[width=0.5\textwidth]{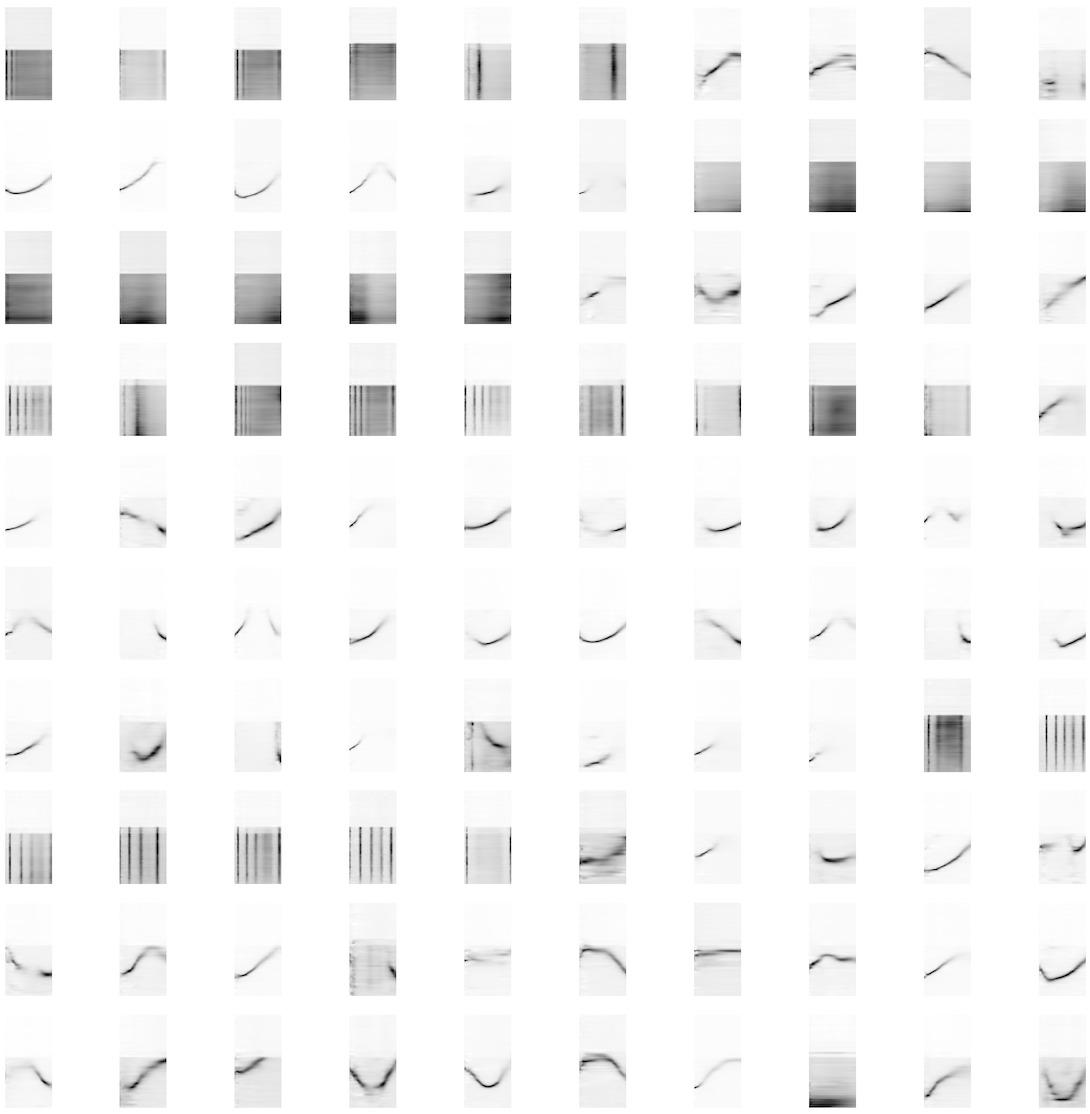}
  \caption{Some reconstructions of dolphin signals. 
    One can clearly see whistles and echolocation sounds. The
    reconstructed whistles include up and down sweeps as well as turning points.
    The echolocation reconstructions are visible as one or more vertical lines.
    Most of the burst pulse sounds are not reconstructed sharply since the stereotypical
    harmonics are not present.
  }
  \label{fig:ae_predict}
\end{figure}

In another test, we visually inspect the reconstructions of several windows using the autoencoder.
As one can see in Fig. \ref{fig:ae_predict} the reconstructions are indeed recognizable as dolphin signals.
Clicks and whistles are especially visible, while burst pulses are often smeared.

\begin{figure}[ht]
  \centering
  \includegraphics[width=0.5\textwidth]{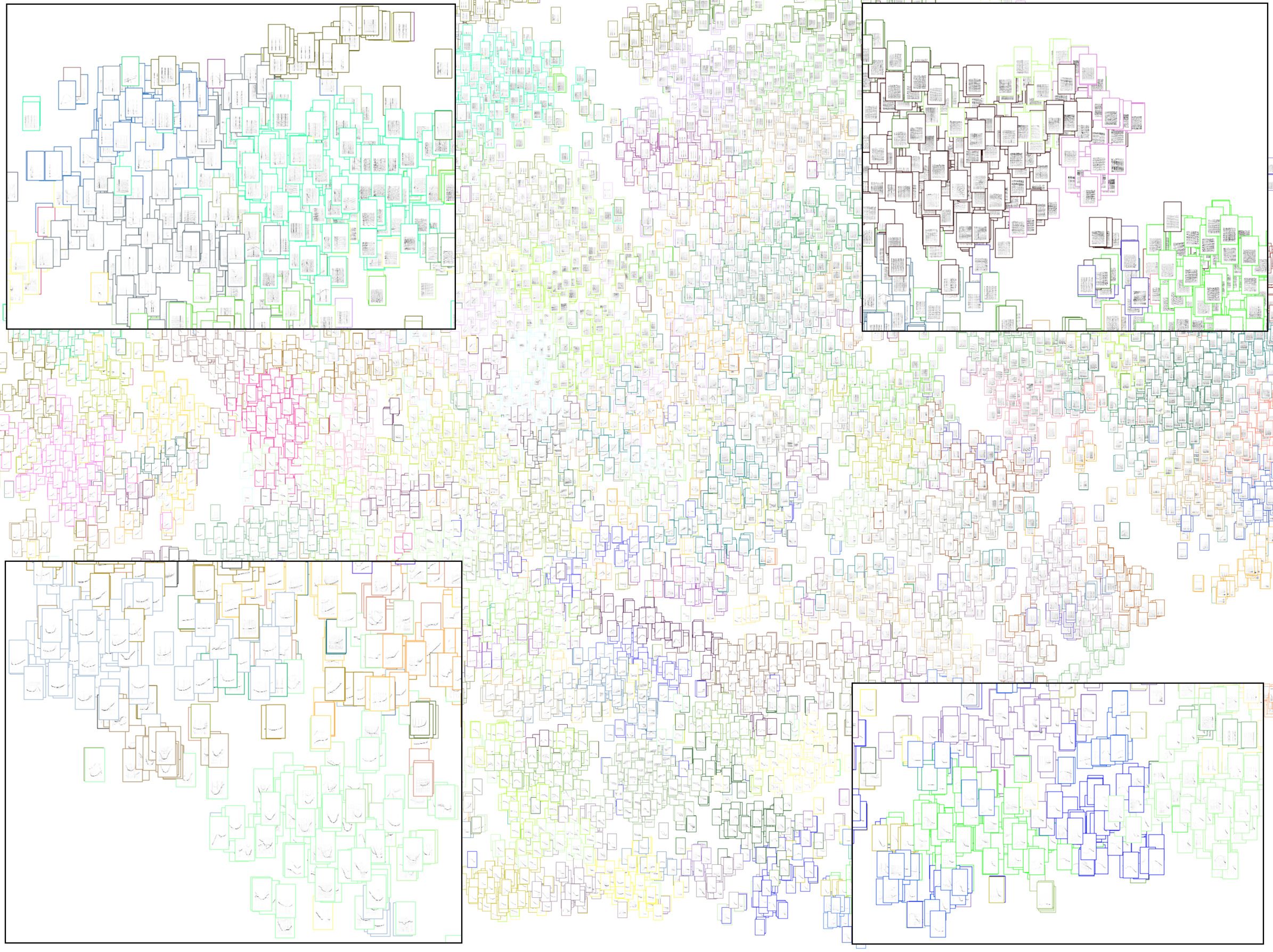}
  \caption{Embedding of the spectrogram windows from the train set.
    The position is determined by the T-SNE projection of the embeddings.
    With colors indicating the cluster, in the background
    we plotted the complete T-SNE map and in the foreground, we zoomed into the four corners.
    Best viewed in color. For details please enlarge.}
  \label{fig:train_embedding}
\end{figure}

\begin{figure}[ht]
  \centering
  \includegraphics[width=0.5\textwidth]{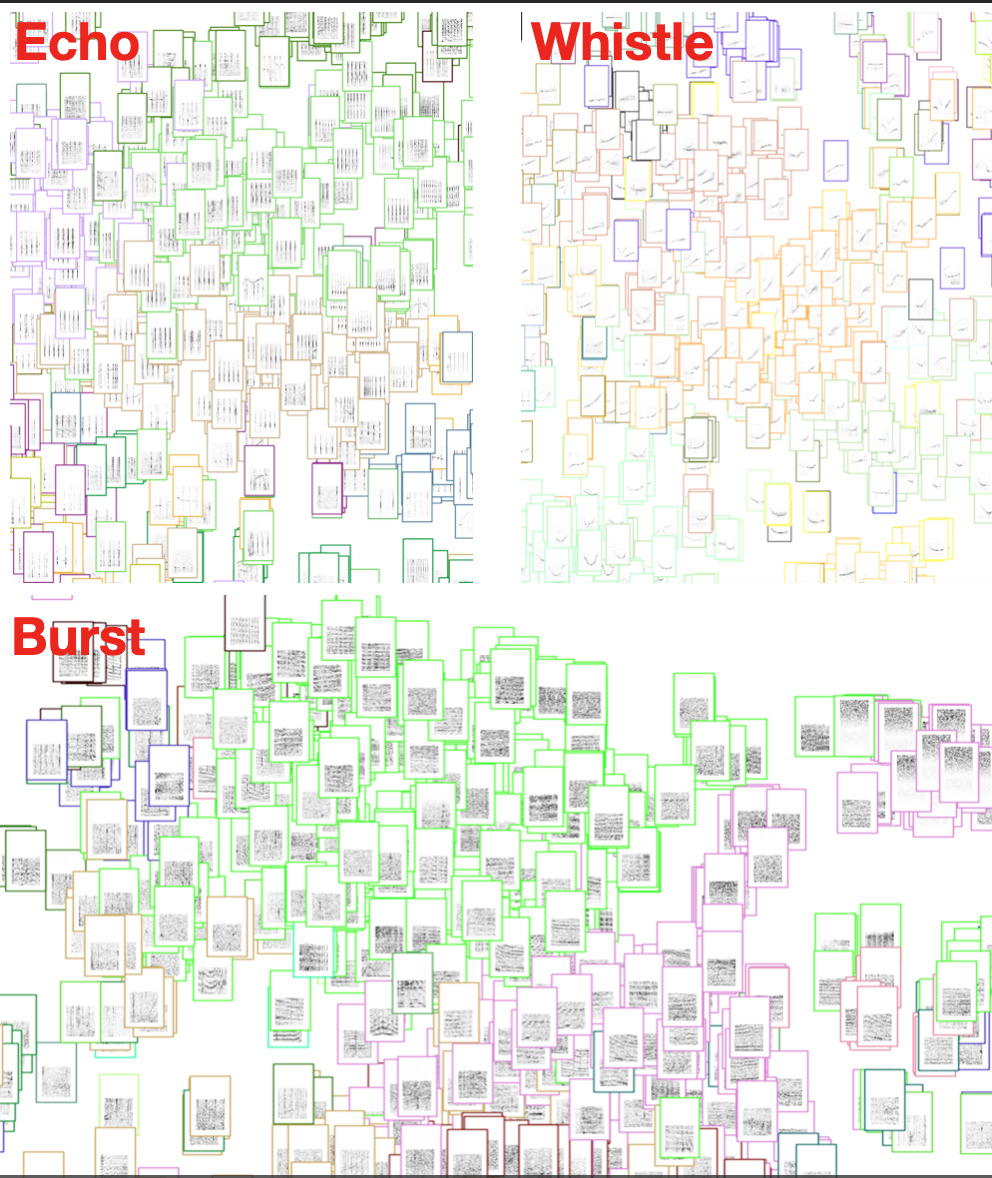}
  \caption{Elarged version of the training set embedding. Signals densely cluster by type and also shape}
  \label{fig:train_embedding_zoom}
\end{figure}

We then embed all of the training data and cluster the resulting vectors using k-means. We use 100 clusters
and initialize the clustering using k-means++ \cite{b17}. We also restrict k-means to 1024 iterations. In order to
visualize the clustering, we project the embeddings into 2D using T-SNE \cite{b11}. The results are plotted in
Figure \ref{fig:train_embedding}. A magnified version is shown in Figure \ref{fig:train_embedding_zoom}.

\subsection*{Signal Detection}
In this experiment, we desire to show that the basic model can be adapted
to the signal detection task. Therefore, we split the signal detection
dataset into 60\% training data and 40\% testing data with each set containing data from all years. We employ a
standard neural network for binary classification.
In this experiment, we add three dense layers on top of the encoder's output.
First, we batch normalize the output of the encoder and then add two dense layers (64 neurons and 32 neurons).
Before adding the classification layer we add one dropout layer with a dropout rate of 0.5.
The last layer has one output neuron with a sigmoid activation.
During training, we use a batch size of 10 instances and train for 25 epochs using an ADAM optimizer.
The results are shown in Table I. As one can see, we achieve an accuracy of 96\% on the test set.
In the experiment, we do not fix the layers of the encoder and initialize the weights with the encoder from the unsupervised experiment.

\begin{table}[ht]
\centering
\begin{tabular}{|c | c | c|} 
 \hline
 truth / prediction & dolphin & noise \\
 \hline\hline
 dolphin & 83 & 10 \\ 
 \hline
 noise   &  5 & 347 \\
 \hline
\end{tabular}
\label{tab:signaldetect}
\caption{The confusion matrix for the signal detection experiment.}
\end{table}

\subsection*{Signal Type Classification}
In another experiment, we adapt the model to perform
dolphin signal type classification.
We classify the dolphin signals into the four categories
contained in the dataset. Again we use a 60 / 40 split
from all years, and we use the same architecture and
training as in the signal detection experiment, except we adjust the last
layer to have four neurons with a softmax activation.
As can be seen in Table II,  we achieve an 85\% accuracy in this experiment.

\begin{table}[ht]
\centering
\begin{tabular}{|c | c | c| c| c|} 
 \hline
 truth / prediction & noise & echo & burst & whistle \\
 \hline\hline
 noise   & 291 & 27 & 15  & 1 \\ 
 \hline
 echo    &  35 & 434 & 9  & 7 \\
 \hline
 burst   &  38 & 30 & 207 & 8 \\
 \hline
 whistle &   8 & 12 &   5 & 181\\
 \hline
\end{tabular}
\caption{The confusion matrix for the signal classification experiment.}
\label{tab:signaltype}
\end{table}

\subsection*{Mining a year of data}
\begin{figure}[ht]
  \centering
  \includegraphics[width=0.5\textwidth]{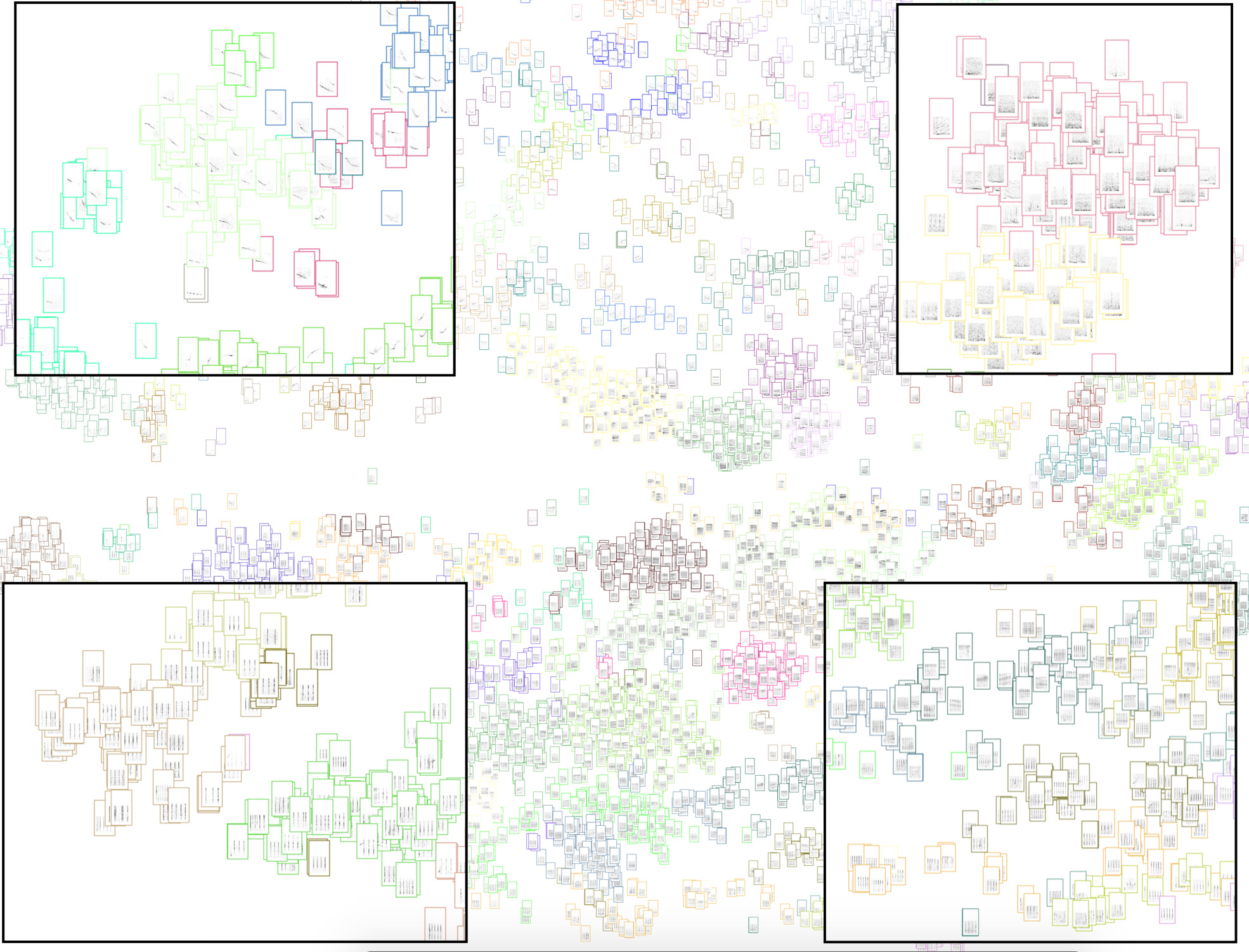}
  \caption{Embedding of the spectrogram windows from the test set. The position is determined
    by the T-SNE projection of the embeddings. The colors indicate the cluster, in the background we plotted the complete
    T-SNE map and in the foreground we zoomed into the four corners.
    Best viewed in color. For details please enlarge.}
  \label{fig:2011embedding}
\end{figure}

\begin{figure}[ht]
  \centering
  \includegraphics[width=0.5\textwidth]{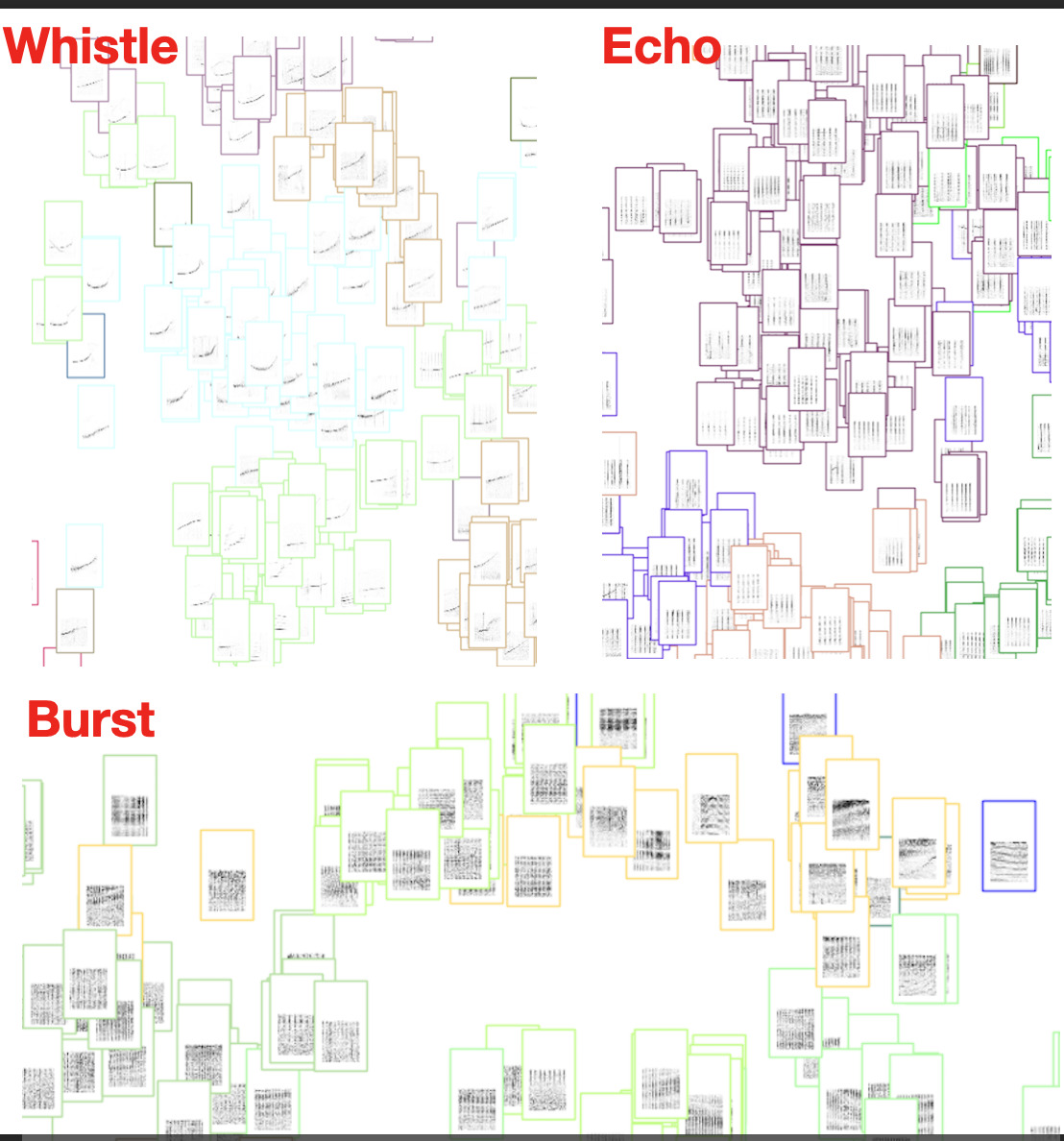}
  \caption{Elarged version of the training set embedding. Signals densely cluster by type and also shape.}
  \label{fig:2011embedding_zoom}
\end{figure}

In our final experiment, we use the trained encoder and the signal detector to extract and cluster patterns
from the whole year of data from 2011. We extract windows classified as dolphin communication using the signal detector.
We then embed the windows using the encoder and cluster the resulting embeddings into $100$ clusters using the
same method as in the previous clustering experiment. We visualize the result in the same way.
After clustering, we estimate the silhouette coefficient for each window. The silhouette coefficient
measures how windows are clustered with similar samples \cite{b23}. We then filter all samples with a coefficient
lower than the medium coefficient across all samples. The results are shown in Fig. \ref{fig:2011embedding}.
In this case, we only plot 25\% of the instances for better visibility.
A zoomed in version is shown in Figure  \ref{fig:2011embedding_zoom}.

\begin{figure}[ht]
  \centering
  \includegraphics[width=0.5\textwidth]{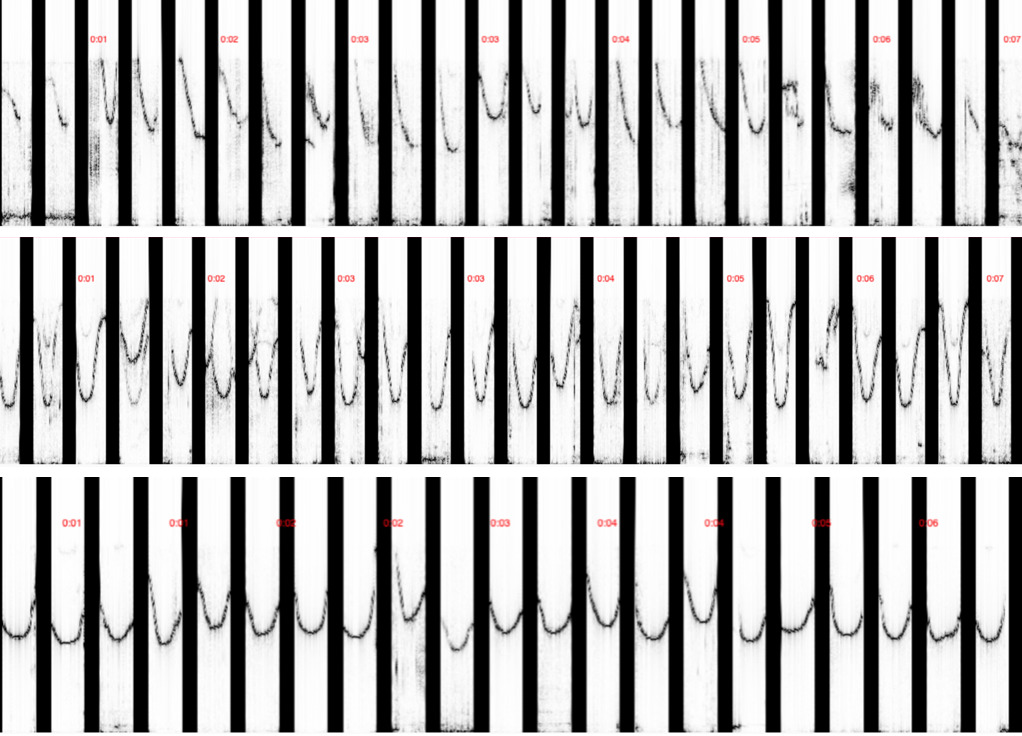}
  \caption{
    A cluster showing spectrograms of a whistle warped in time and shifted in frequency.
    Especially in the middle row, we see several distortions of the signal including overlapping
    signals. 
  }
  \label{fig:warping}
\end{figure}

We inspect all clusters visually as well. Therefore, we export each cluster into a single wav file. The audio in each window 
is concatenated with a short gap of silence (see Fig. \ref{fig:warping}). We then inspect each spectrogram and note how many clusters have the same type
and how many silence clusters still made it through the silence detector. We found that 14 clusters are still mostly noise
resulting in an $86\%$ accuracy for the silence detector. We also found four clusters of mixed type (mix of whistles, burst sounds and echolocation).

In the example in Figure \ref{fig:warping} we see three clusters with slight time warps and shifts in frequency.

\section*{Discussion}
In our experiments, we showed that our model is indeed able to learn an embedding for audible dolphin signals in an unsupervised manner.
We showed that the embedding can be trained using an autoencoder that reconstructs dolphin signals. Fig. \ref{fig:embedder_filters}
shows the first layer's convolutional kernels. 
Nearly all of these kernels look like spectrogram patches from dolphin communication which can be interpreted as
evidence for successful training. In the same way, Fig. \ref{fig:ae_predict} presents successful reconstructions of the autoencoder as further evidence.
While we are able to successfully reconstruct whistles and echolocation sounds, the burst pulses blur more.
Furthermore, 
by visual inspection, the clusters in the 2D projection seem tight, meaning that the embedding is able to model patterns in dolphin communication. We also showed that the embedder
can be used for transfer learning. The signal detector and the type classifier achieve high accuracy on their test sets despite the significantly smaller training
data. Because labeling of dolphin signals is cumbersome for marine mammalogists it is expected that more directed efforts towards dolphin communication in
specific behavioral contexts will produce smaller datasets. However, our transfer learning results indicate that the model can be adjusted with little data
to other tasks.
In the final experiment, we ran the signal detector and the embedder in a more realistic scenario. We showed that the signal detector performs well across data
recorded throughout a whole year. Furthermore, we showed that the embedding of the detected regions produces clean clusters. In total, we think that we trained a successful feature
extractor which will be the basis of our future research. 

\section*{Future Work}
Our experiments show that the model is indeed an effective model of short term fixed-size windows of dolphin communication. In the future, we plan to investigate several methods for sequence analysis, in contrast to the fixed sized windows utilized in our approach. Sliding the encoder across the spectrogram together with the silence detector
will create variable length embedding sequences of dolphin communication bounded by silence. We aim to replace the k-means clustering of single embedding vectors
with agglomerative clustering of sequences of embedding vectors using the dynamic time warping distance. Furthermore, we aim to create multiple sequence alignments of
dolphin communication. We hope that the visualization of aligned spectrograms
will ease the comparison of longer sequences of dolphin communication.
Finally, we seek to answer questions about the structure of dolphin communication. One question marine mammalogists have is 
whether dolphin communication displays a similar structure to human communication. One idea is to use recursive neural networks \cite{b5} on top of sequences of embedding vectors in order
to find structural patterns.

\section*{Conclusion}
We proposed an autoencoder constructed from convolutional and recurrent layers in order to construct an embedding of short windows of audible dolphin communication.
Our architecture is inspired by encoder-decoder models in natural language processing. The encoder transforms short spectrogram windows into a single encoding vector.
The decoder reconstructs the complete spectrogram from just the embedding vector. In a series of experiments, we showed the effectiveness of the embedding.
First, we visualized the first layer's convolutions and found that the convolutional kernels seem to pick up on dolphin signals. When plotting the decoder's reconstructions
we clearly saw that we are able to encode enough information in the embedding vector to reconstruct the signals. When clustering the embedded spectrograms we saw that the clusters seem distinct as well. We also showed the model's performance on two transfer learning tasks. In a signal detection experiment and a type classification experiment, the retrained
model showed high accuracies. When running the trained models on a whole year of data we found good clustering and signal detection performance.

\end{document}